\documentclass{jpsj-suppl}
\usepackage{txfonts} 

\def\pandas{\ifmmode\mathrm{\overline{\textsc{P}}{\textsc{anda}}}
            \else$\mathrm{\overline{\textsc{P}}{\textsc{anda}}}$\fi}
\def\panda{\ifmmode\mathrm{\overline{{P}}{{ANDA}}}
            \else$\mathrm{\overline{{P}}{{ANDA}}}$\fi}

\def\GeV{\,GeV}
\def\kevc1{\ifmmode\mathrm{\ keV/{\mit c}}
          \else$\mathrm{\ keV/{\mit c}}$\fi}

\def\MeVc1{\,MeV/{\mit c}}

\def\mevc1{\ifmmode\mathrm{\ MeV/{\mit c}}
          \else$\mathrm{\ MeV/{\mit c}}$\fi}
\def\gevc1{\ifmmode\mathrm{\ GeV/{\mit c}}
          \else$\mathrm{\ GeV/{\mit c}}$\fi}
\def\GeVc1{\ifmmode\mathrm{\ GeV/{\mit c}}
          \else$\mathrm{\ GeV/{\mit c}}$\fi}
\def\kevc2{\ifmmode\mathrm{\ keV/{\mit c}^2}
          \else$\mathrm{\ keV/{\mit c}^2}$\fi}
\def\Mevc2{\ifmmode\mathrm{\ MeV/{\mit c}^2}
          \else$\mathrm{\ MeV/{\mit c}^2}$\fi}
\def\Gevc2{\ifmmode\mathrm{\ GeV/{\mit c}^2}
          \else$\mathrm{\ GeV/{\mit c}^2}$\fi}
\def\Gev2c2{\ifmmode\mathrm{\ GeV^2/{\mit c}^2}
          \else$\mathrm{\ GeV^2/{\mit c}^2}$\fi}
\def\Pgp{\ifmmode\mathrm{p}
         \else$\mathrm{p}$\fi}
\def\Pagp{\ifmmode\mathrm{\overline{p}}
         \else$\mathrm{\overline{p}}$\fi}
\def\Pgn{\ifmmode\mathrm{n}
         \else$\mathrm{n}$\fi}
\def\Pagn{\ifmmode\mathrm{\overline{n}}
         \else$\mathrm{\overline{n}}$\fi}
\def\Pp{\ifmmode\mathrm{p}
         \else$\mathrm{p}$\fi}
\def\Pap{\ifmmode\mathrm{\overline{p}}
         \else$\mathrm{\overline{p}}$\fi}

\def\Pn{\ifmmode\mathrm{n}
         \else$\mathrm{n}$\fi}
\def\Pan{\ifmmode\mathrm{\overline{n}}
         \else$\mathrm{\overline{p}}$\fi}
\def\Py{\ifmmode\mathrm{Y}
         \else$\mathrm{Y}$\fi}
\def\Pay{\ifmmode\mathrm{\overline{Y}}
         \else$\mathrm{\overline{Y}}$\fi}

\def\PgL{\ifmmode\mathrm{\Lambda}
          \else$\mathrm{\Lambda}$\fi}
\def\PagL{\ifmmode\mathrm{\overline{\Lambda}}
            \else$\mathrm{\overline{\Lambda}}$\fi}
\def\PgS{\ifmmode\mathrm{\Sigma}
          \else$\mathrm{\Sigma}$\fi}
\def\PagS{\ifmmode\mathrm{\overline{\Sigma}}
            \else$\mathrm{\overline{\Sigma}}$\fi}
\def\PgSp{\ifmmode\mathrm{\Sigma^+}
          \else$\mathrm{\Sigma^+}$\fi}
\def\PagSp{\ifmmode\mathrm{\overline{\Sigma^+}}
            \else$\mathrm{\overline{\Sigma^+}}$\fi}
\def\PgSm{\ifmmode\mathrm{\Sigma^-}
          \else$\mathrm{\Sigma^-}$\fi}
\def\PagSm{\ifmmode\mathrm{\overline{\Sigma^-}}
            \else$\mathrm{\overline{\Sigma^-}}$\fi}
\def\PgSn{\ifmmode\mathrm{{\Sigma}^0}
            \else$\mathrm{{\Sigma}^0}$\fi}
\def\PagSn{\ifmmode\mathrm{\overline{\Sigma}^0}
            \else$\mathrm{\overline{\Sigma}^0}$\fi}
\def\PgX{\ifmmode\mathrm{\Xi}
          \else$\mathrm{\Xi}$\fi}
\def\PagX{\ifmmode\mathrm{\overline{\Xi}}
            \else$\mathrm{\overline{\Xi}}$\fi}
\def\PgXm{\ifmmode\mathrm{\Xi^-}
          \else$\mathrm{\Xi^-}$\fi}
\def\PagXm{\ifmmode\mathrm{\overline{\Xi^-}}
            \else$\mathrm{\overline{\Xi^-}}$\fi}
\def\PagXp{\ifmmode\mathrm{\overline{\Xi}}\mathrm{^+}
            \else$\mathrm{\overline{\Xi}}\mathrm{^+}$\fi}
 \def\PagXn{\ifmmode\mathrm{\overline{\Xi}^0}
            \else$\mathrm{\overline{\Xi}^0}$\fi}
\def\PgO{\ifmmode\mathrm{\Omega}
          \else$\mathrm{\Omega}$\fi}
\def\PagO{\ifmmode\mathrm{\overline{\Omega}}
            \else$\mathrm{\overline{\Omega}}$\fi}
\def\PgOm{\ifmmode\mathrm{\Omega^-}
          \else$\mathrm{\Omega^-}$\fi}
\def\PagOm{\ifmmode\mathrm{\overline{\Omega^-}}
            \else$\mathrm{\overline{\Omega^-}}$\fi}
\def\PgOp{\ifmmode\mathrm{\Omega^+}
          \else$\mathrm{\Omega^+}$\fi}
\def\PagOp{\ifmmode\mathrm{{\overline{\Omega}}\mathrm{^+}}
            \else$\mathrm{\overline{\Omega}}\mathrm{^+}$\fi}

\def\PgLc{\ifmmode\mathrm{\Lambda_c}
          \else$\mathrm{\Lambda_c}$\fi}
\def\PagLc{\ifmmode\mathrm{\overline{\Lambda}_c}
            \else$\mathrm{\overline{\Lambda}_c}$\fi}

\def\PgD{\ifmmode\mathrm{D}
          \else$\mathrm{D}$\fi}
\def\PagD{\ifmmode\mathrm{\overline{D}}
            \else$\mathrm{\overline{D}}$\fi}

\def\PgPi{\ifmmode\mathrm{\pi}
          \else$\mathrm{\pi}$\fi}
\def\PagPi{\ifmmode\mathrm{\overline{\pi}}
            \else$\mathrm{\overline{\pi}}$\fi}

\title{Many Facets of Strangeness Nuclear Physics with Stored Antiprotons}

\author{Josef \textsc{Pochodzalla}$^{1,2}$,
Sebastian \textsc{Bleser}$^{2}$,
Alicia \textsc{Sanchez Lorente}$^{2}$,
Marta \textsc{Mart\`{\i}nez Rojo}$^{2}$,
Marcell \textsc{Steinen}$^{2}$
for the $\panda$ Collaboration$^{3}$\\
and
J\"urgen \textsc{Gerl}$^{4}$,
Jasmina \textsc{Kojouharova}$^{5}$,
Ivan \textsc{Kojouharov}$^{4}$
}

\inst{
$^{1}${Institut f\"ur Kernphysik, Johannes Gutenberg-Universit\"at, 55099 Mainz, Germany}\\
$^{2}${Helmholtz Institute Mainz, 55099 Mainz, Germany}\\
$^{3}${https://panda.gsi.de/download/panda\_authors.pdf}\\
$^{4}${GSI Helmholtzzentrum f\"ur Schwerionenforschung, 64291 Darmstadt, Germany}\\
$^{5}${Technische Hochschule Mittelhessen, 61169 Friedberg, Germany}
}

\email{pochodza@uni-mainz.de}

\recdate{January 11, 2016}

\abst{Stored antiprotons beams in the GeV range represent a unparalleled factory for hyperon-antihyperon pairs.
Their outstanding large production probability in antiproton collisions will open the floodgates for a series
of new studies of strange hadronic systems with unprecedented precision. The behavior of hyperons and -- for the first time -- of antihyperons in nuclear systems can be studied under well controlled conditions. The exclusive production of $\PgL\PagL$ and $\PgSm\PagL$ pairs in antiproton-nucleus interactions probe the neutron and proton distribution in the nuclear periphery and will help to sample the neutron skin. For the first time, high resolution $\gamma$-spectroscopy of doubly strange nuclei will be performed, thus complementing measurements of ground state decays of double hypernuclei with mesons beams at J-PARC or possible decays of particle unstable hypernuclei in heavy ion reactions. High resolution spectroscopy of multistrange $\Xi$-atoms are feasible and even the production of $\PgOm$-atoms will be within reach. The latter might open the door to the $|s|$=3 world in strangeness nuclear physics, by the study of the hadronic $\PgOm$-nucleus interaction and the very first measurement of a spectroscopic quadrupole moment of a baryon which will be a benchmark test for our understanding of hadron structure.}

\kword{strangeness, antibaryons, hypernuclei, hyperatoms}

\begin{document}
\maketitle

\section{Introduction}
\label{sec:intro}
Strangeness may exist in our present universe, namely in the core of compact stellar objects like neutron stars. These stars are unique because all four known fundamental interactions contribute to their properties. These objects can give information on the state of matter in extreme regimes and may eventually become an extremely important reference point for alternative theories of gravity once the other interactions are well under control. Indeed, the recent observation of massive neutron stars \cite{Dem10,Ant13} and the expected appearance of hyperons at about two times nuclear density remains an unresolved mystery (“hyperon puzzle”). At present, our incomplete understanding of the underlying baryon-baryon and even more subtle multi-body interactions in baryonic systems limits probably our knowledge of the flavor composition of neutron stars.

\begin{figure}
 \centering
 \includegraphics[width=0.99\textwidth]{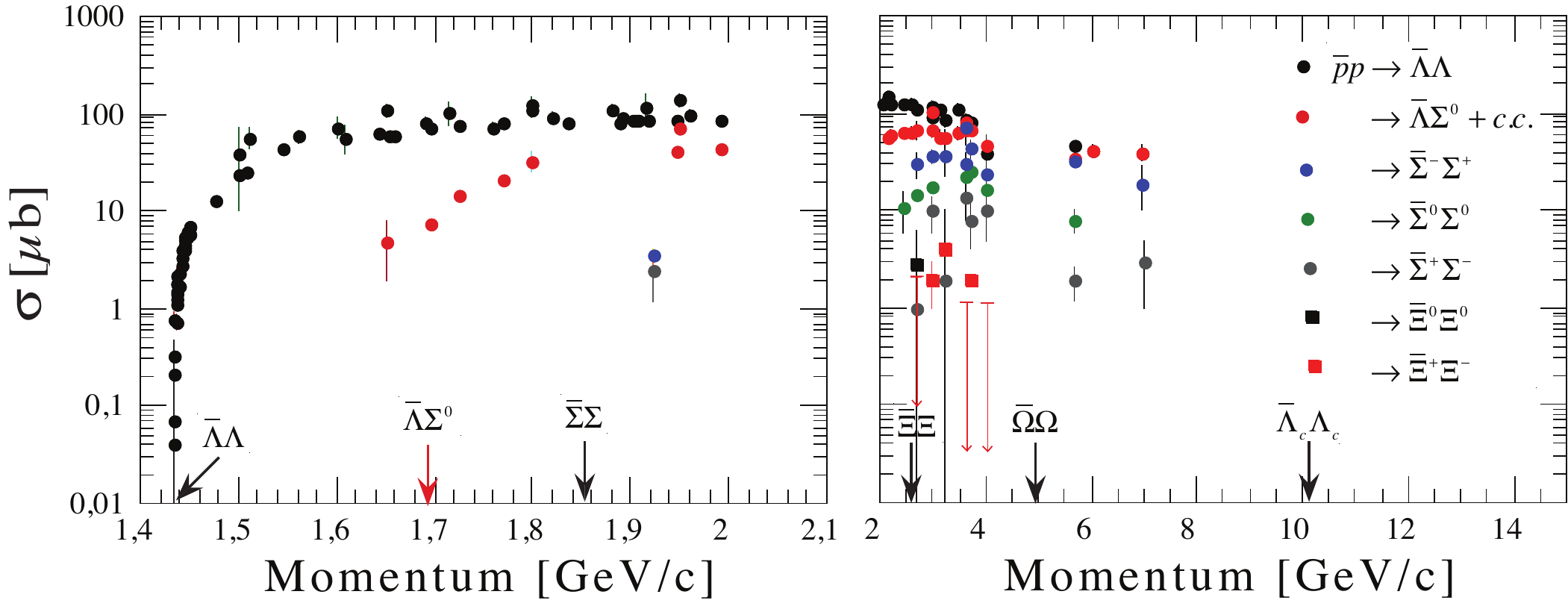} \hfill
 \caption{Total cross sections for reactions  $\Pagp\Pgp$ $\rightarrow$ $\Py\Pay$ in the momentum range of the HESR.
 The upper limits in red refer to the $\PagXp\PgXm$-channel. The arrows pointing to the momentum axis indicate the
 threshold momenta for the different hyperon
families (from Ref.~\cite{Pan09}).}
 \label{fig:xsec}
\end{figure}

Bound strange systems - hypernuclei as well as hyperatoms - represent unique laboratories for multi-baryon interactions with strangeness.  Antiproton-nucleon annihilations constitute the most effective way to produce low momentum hyperons and antihyperons under controlled kinematic conditions which is a prerequisite for the formation of bound hyperonic systems. Combined with large cross sections for the production of associated hyperon-antihyperon pairs (see Fig. \ref{fig:xsec}), antiprotons circulating in an storage ring are ideally suited for exploring strange baryonic systems.

\section{Antihyperons in Nuclear matter}
\label{sec:anti}
The interaction of antibaryons in nuclei provides a unique opportunity to elucidate strong in-medium effects in baryonic systems. Unfortunately, information on antibaryons in nuclei are rather scarce and only for the antiproton the nuclear potential could be constrained by experimental studies. The (Schr\"odinger equivalent) antiproton potential at normal nuclear density turns out to be in the range of $U_{\overline{p}} \simeq -150$MeV, i.e. a factor of approximately four weaker than naively expected from G-parity relations \cite{Lar09}. Gaitanos {\em et al.} \cite{Gai11} suggested that this discrepancy can be traced back to the missing energy dependence of the proton-nucleus optical potential in conventional relativistic mean-field models. The required energy and momentum dependence could be recovered by extending the relativistic hadrodynamics Lagrangian by non-linear derivative interactions \cite{Gai09,Gai11,Gai13} thus also mimicking many-body forces \cite{Gom14}. Considering the important role played e.g. by strange baryons and antibaryons for a quantitative interpretation of high-energy heavy-ion collisions and dense hadronic systems it is clearly mandatory to test these concepts also in the strangeness sector.

Antihyperons annihilate quickly in nuclei and conventional spectroscopic studies of bound systems are not feasible. As a consequence, no experimental information on the nuclear potential of antihyperons exists so far. As suggested recently \cite{Poc08,Poc09,San15}, quantitative information on the antihyperon potentials may be
obtained via exclusive antihyperon-hyperon pair production close to threshold in antiproton-nucleus interactions.
The schematic calculations of Ref. \cite{Poc08,Poc09} revealed significant sensitivities of the transverse momentum asymmetry $\alpha_{T}$ which is defined in terms of the transverse momenta of the coincident particles
\begin{equation}
\alpha_{T}=\frac{p_{T}(\PgL)-p_{T}(\PagL)}{p_{T}(\PgL)+p_{T}(\PagL)}
\label{eq:00}
\end{equation}
to the depth of the antihyperon potential. In order to go beyond the schematic calculations presented in Refs.
\cite{Poc08,Poc09} and to include simultaneously secondary deflection and absorption effects, we recently analyzed \cite{San15} more realistic calculations of this new observable with the Giessen Boltzmann-Uehling-Uhlenbeck transport model (GiBUU, Release 1.5) \cite{GiBUU}. In order to explore the sensitivity of the transverse momentum asymmetry on the depth of the $\PagL$-potential calculations where the default antihyperon potentials \cite{Lar12} were scaled, leaving all other input parameters of the model unchanged.

Besides hydrogen isotopes, also nobel gases can be used by default as nuclear targets by the $\panda$ experiment.
We, therefore, studied the exclusive reaction \Pap+$^{20}$Ne$\rightarrow$ \PagL\PgL\ at beam energies of 0.85\GeV\ and 1\GeV. These energies correspond to antiproton momenta of 1.522\GeVc1 and 1.696\GeVc1, respectively.
At 0.85\GeV\ the \PagS\PgL\ and \PgS\PagL\ channels are not accessible and also the production of a pion in addition to a $\PgL\PagL$ pair can be neglected. The higher energy of 1\GeV\ lies above the $\PagS\PgL$- threshold and makes also the \Pap\Pn\ $\rightarrow\PagL\PgSm$ and \Pap\Pp\ $\rightarrow$\PagL\PgSn\ as well as their charge conjugate channels accessible.
\begin{figure}[tb]
\includegraphics[width=0.50\textwidth]{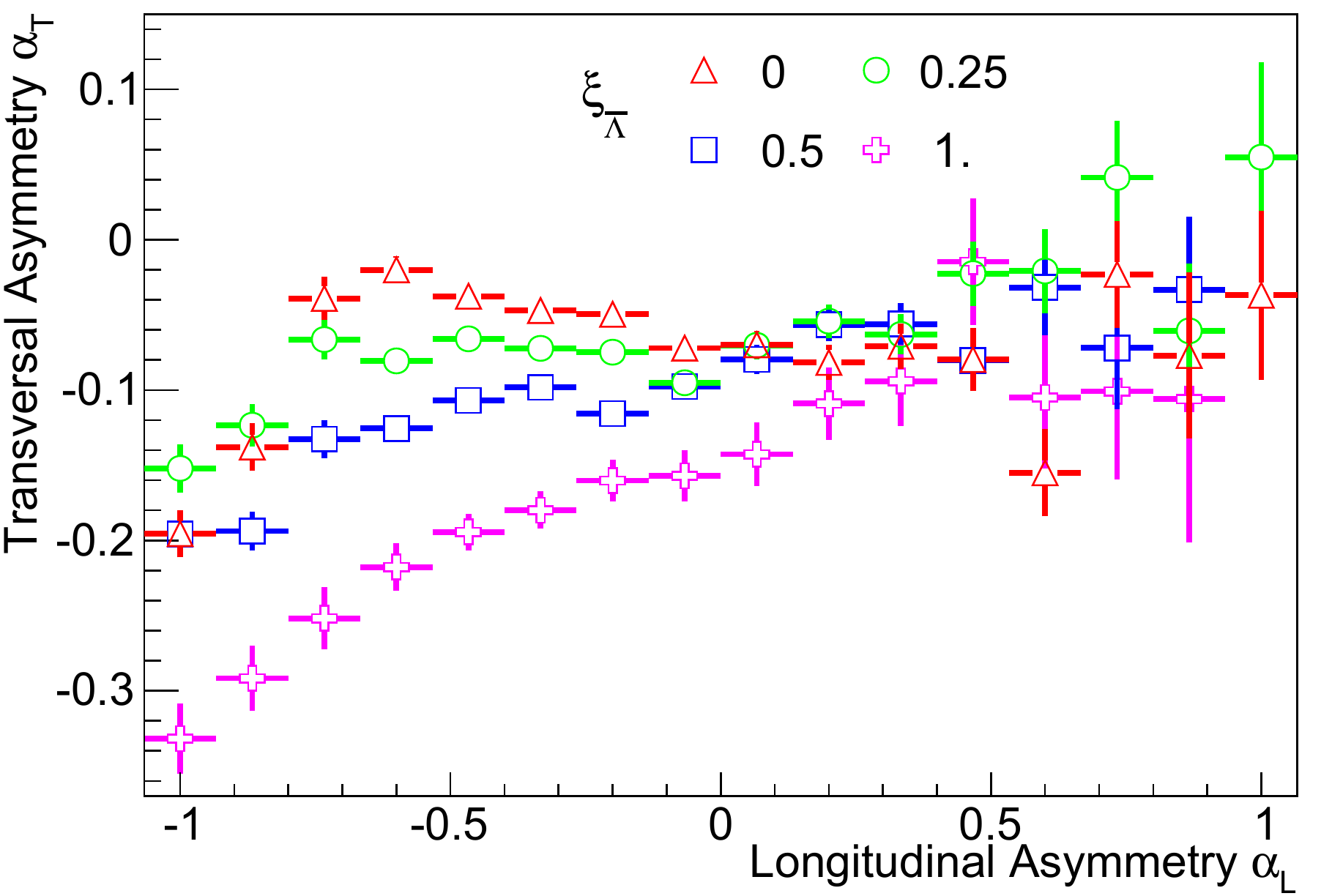}
\includegraphics[width=0.48\textwidth]{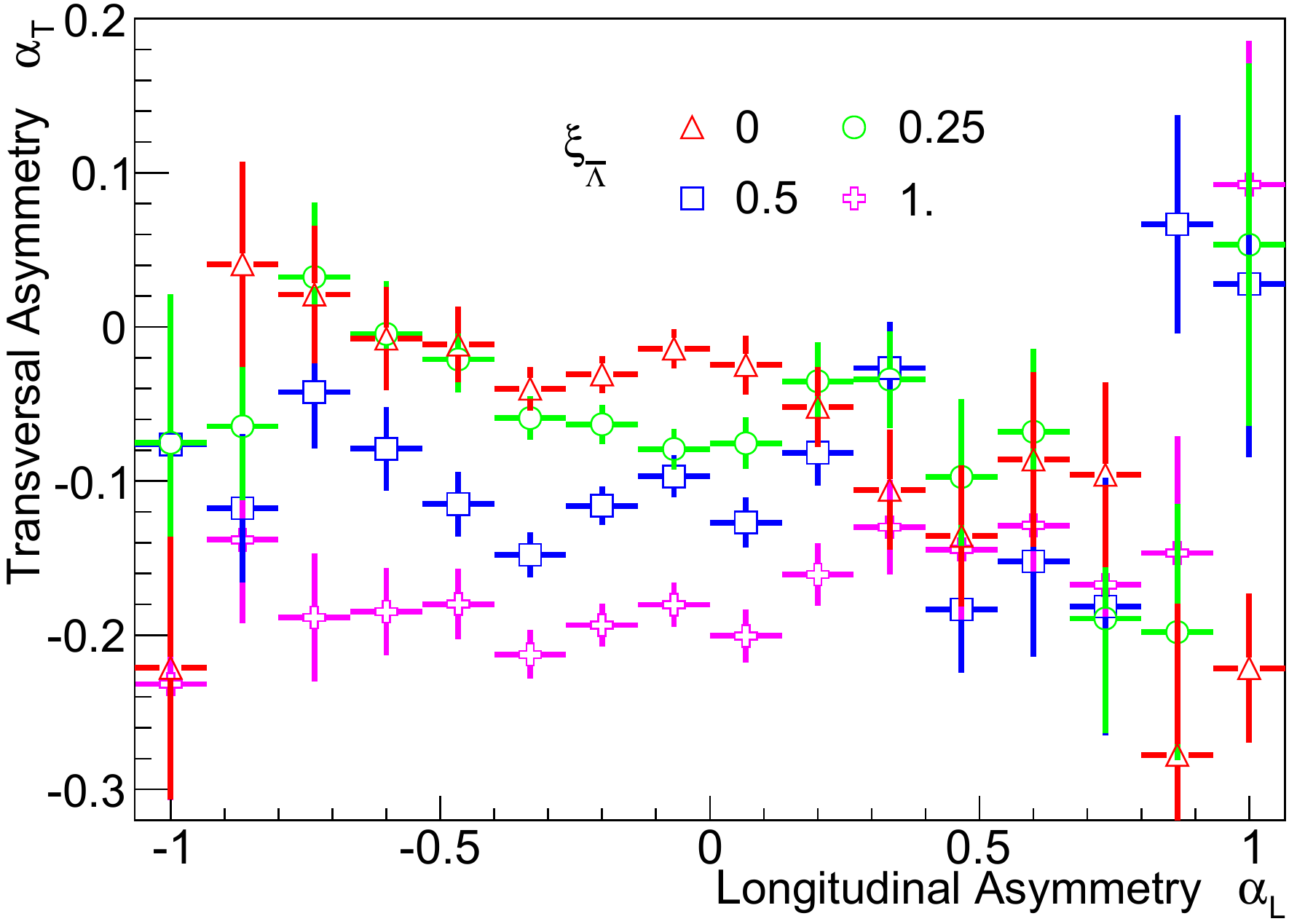}
\caption{Average transverse momentum asymmetry as a function of the
longitudinal momentum asymmetry for $\PgL\PagL$ pairs (left; from Ref. \cite{San15}) and $\PgSm\PagL$ pairs (right) produced
exclusively in 1\GeV\ $\Pap$-$^{20}$Ne interactions. The different symbols show the GiBUU predictions for different scaling factors $\xi_{\PagL}$ of the $\PagL$-potentials.}
\label{fig:asym}
\end{figure}
Fig.~\ref{fig:asym} shows the GiBUU prediction for the average transverse asymmetry $\alpha_{T}$ (Eq. \ref{eq:00})
plotted as a function of the longitudinal momentum
asymmetry $\alpha_{L}$ which is defined for each event as
\begin{equation}
\noindent
\alpha_{L}=\frac{p_{L}(\PgL)-p_{L}(\PagL)}{p_{L}(\PgL)+p_{L}(\PagL)}.
\label{eq:02}
\end{equation}
Both, for $\PgL\PagL$ pairs (left) and $\PgSm\PagL$ pairs (right) a remarkable sensitivity of $\alpha_{T}$ on the
scaling factors  $\xi_{\PagL}$ of the $\PagL$-potential is found at negative values of $\alpha_{L}$.
In Ref.~\cite{San15} it was demonstrated that the substantial sensitivity of transverse momentum correlations of coincident $\PgL\PagL$ pairs to the assumed depth of the $\PagL$-potential is strongly related to the rescattering process of the hyperons and antihyperons within the target nucleus.
For positive values of $\alpha_{L}$ where the antihyperon is emitted backward with respect
to $\PgL$-particle, the statistics is too low to draw quantitative conclusions in the present simulation.

As discussed in more detail in section~\ref{sec:skin}, $\PgSm\PagL$ pairs are produced in $\Pap$-$\Pn$ interactions. As a consequence a comparison of $\PgL\PagL$ and $\PgSm\PagL$-correlations in neutron-rich nuclei might help to explore the isospin dependence of the $\PagL$-potential in nuclear matter in future. Also $\PgSm\PagL$-momentum correlations show a strong sensitivity to the depth of the antihyperon potential (right part of Fig.~\ref{fig:asym}). However, as already stressed in Ref. \cite{Lar12}, the attractive $\PgSm$-potential adopted in the present GiBUU model is not compatible with experimental data. Clearly, the description of the $\PgSm$-potential in the transport model needs to be improved so that the sensitivities of the momentum correlations shown in the right part of Fig.\,\ref{fig:asym} can be put on firm footing.

\section{Sampling the Neutron Skin}
\label{sec:skin}
\begin{figure}[ b]
\includegraphics[width=0.49\textwidth]{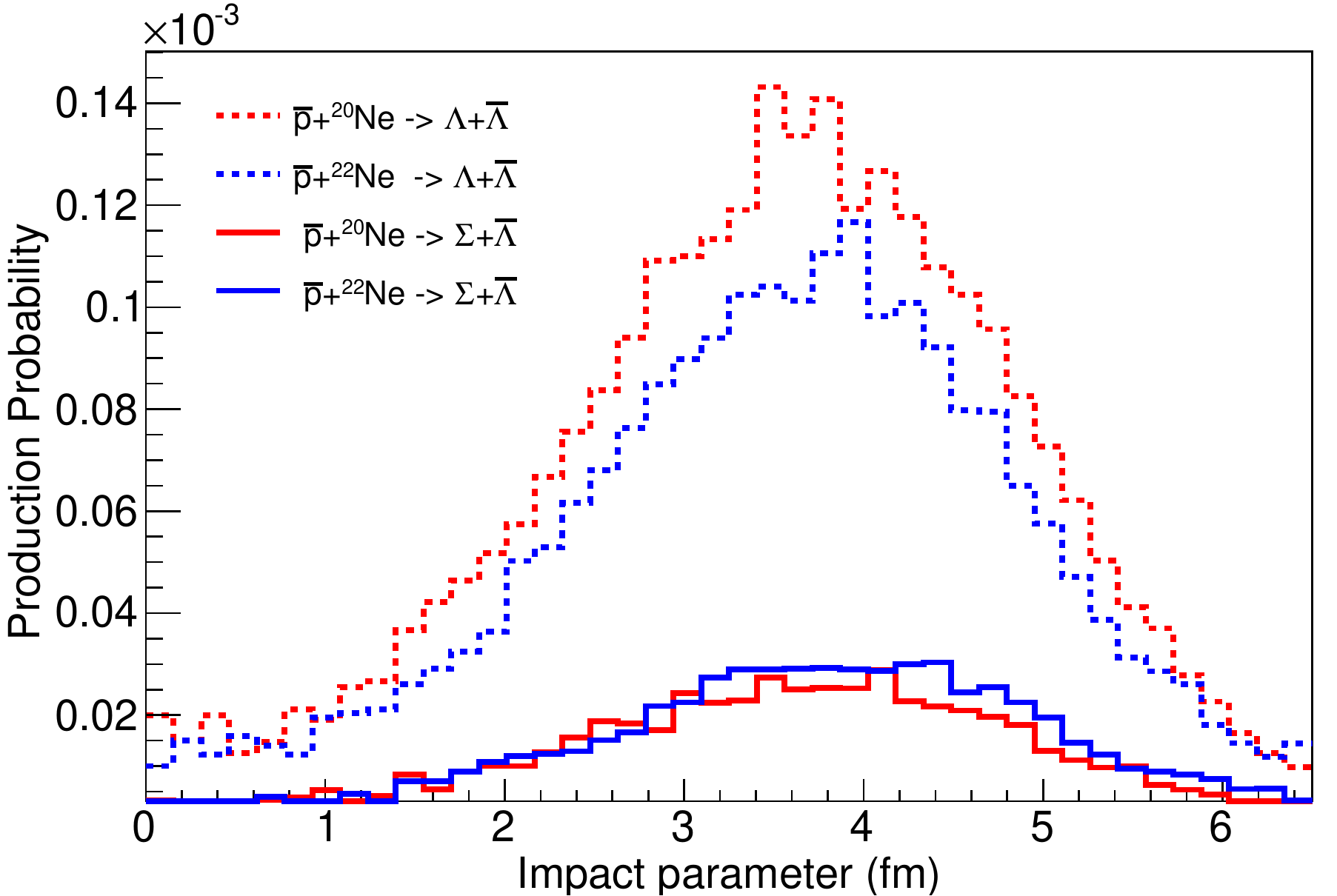}
\includegraphics[width=0.51\textwidth]{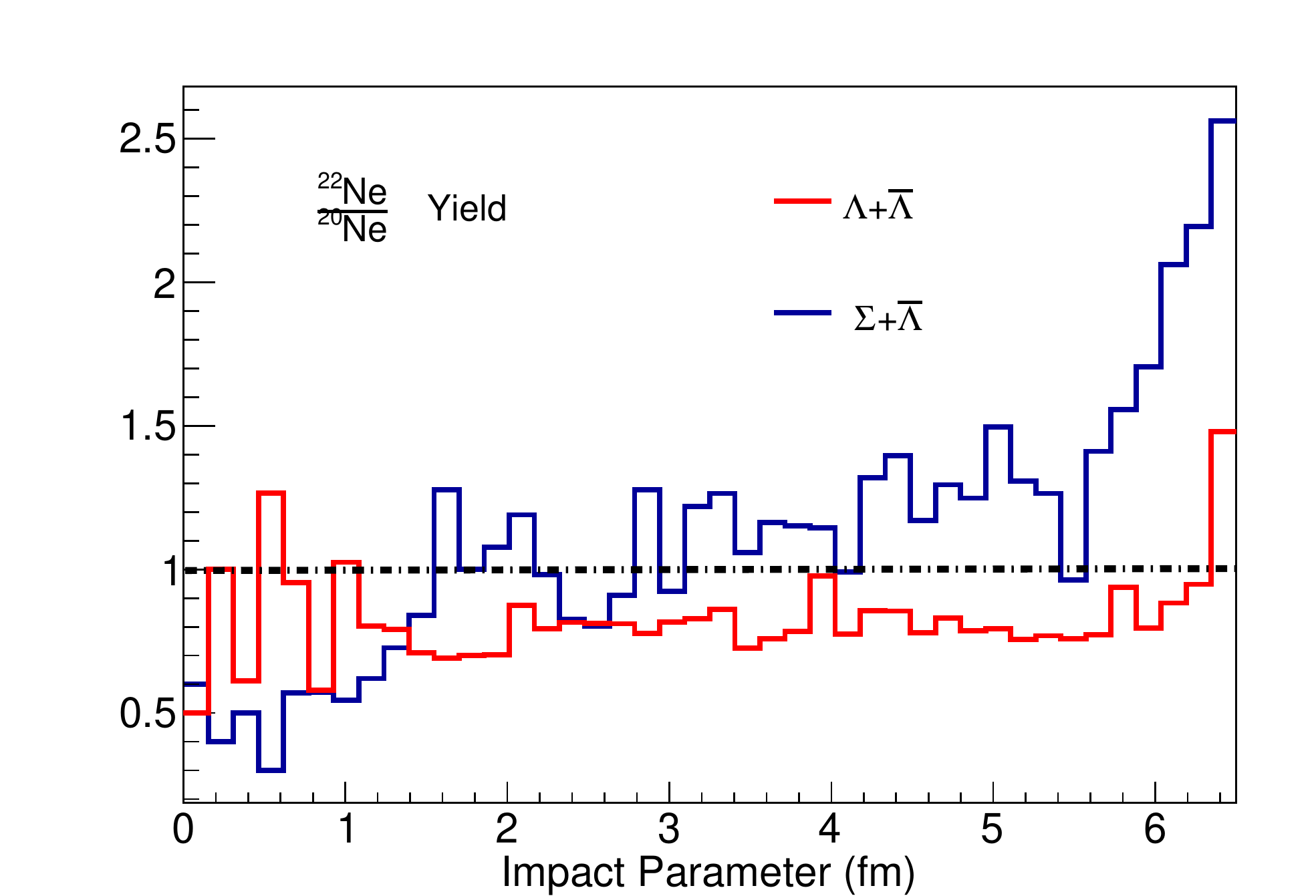}
\caption{Left: Production probability of $\PgL\PagL$ pairs (dashed histograms) and $\PgSm\PagL$ pairs (solid histograms)
produced exclusively in 1\GeV\ $\Pap$-$^{20}$Ne (red histograms)  and $\Pap$-$^{22}$Ne (blue histograms) interactions as a function of the impact parameter. For the $\PagL$-potential a scaling factor of $\xi_{\PagL}$ = 0.25 was used.
Right: Ratio of exclusive $\PgL\PagL$ (red) and $\PgSm\PagL$ pair (blue) production in $^{22}$Ne vs. $^{20}$Ne nuclei.}
\label{fig:skin}
\end{figure}
Conventional atomic nuclei are composed of protons and neutrons. The difference in their spatial distributions represents a sensitive probe for the isospin dependence of the nuclear equation of state and, thus, offers an exciting link to nuclear systems with extreme isospin like dense stellar objects \cite{Hor01}. The proton distributions of nuclei can precisely be determined by the elastic scattering of electrons. For the neutron distribution different tools like dipole polarizability studies \cite{Ros13,Roc15}, pion photoproduction \cite{Tab14}, hadronic scattering and reactions \cite{Dai15} and parity-violating electron scattering \cite{Abr12} are available. In Ref. \cite{Len07} it was already pointed out that in neutron-rich nuclei the outer neutron layer partially shields the absorption on the protons in the interior. With a typical inelastic cross section for antiprotons of 100\,mb and assuming for simplicity spherical neutron and proton distributions with constant density and with $\rho_n$=$\rho_p$, the absorption length of antiprotons in the neutron skin amounts to about 1.2\,fm.
As a consequence of this small distance, the interaction of antiprotons on nuclei are a sensitive probe of the nuclear periphery \cite{Len07,San15}. Furthermore, the exclusive production of hyperon-antihyperon pairs in antiproton-nucleus collisions close to threshold allows to sample the proton and neutron structure separately: while $\PgL\PagL$ pairs are produced in {\Pap}-{\Pp} reactions, $\PgSm\PagL$ pairs signal {\Pap}-{\Pn} collisions.

$^{22}$Ne is besides $^{20}$Ne a conceivable target nucleus for the $\panda$ experiment. The measured charge radii of $^{20}$Ne (r$_c$=2.992$\pm$0.008\,fm) and $^{22}$Ne (r$_c$=2.986$\pm$0.021\,fm) differ only very little \cite{DeV87}. This small variation is also reproduced by Hatree-Fock-Bogoliubov \cite{Gru96} and relativistic mean field calculations \cite{Pan14}. On the other hand according to these calculations, the radius of the neutron distribution of $^{20}$Ne is about 0.2\,fm smaller than the one of $^{22}$Ne \cite{Gru96}.
With the simplifying assumptions mentioned above and considering the case of non-peripheral $\Pap$ - Ne collisions at impact parameters b $<$ r$_c$, typically p$_{abs}$ =16\% of all impinging $\Pap$ will be absorbed within the additional neutron layer of $^{22}$Ne. While the exclusive $\PgSm\PagL$ production in $^{22}$Ne as compared to $^{20}$Ne will be enhanced by a factor (1+p$_{abs}$)$\approx$1.16, the $\PgL\PagL$ pair production in $\Pap$-$\Pp$ reactions will be reduced by a factor (1-p$_{abs}$)$\approx$0.84.

Of course, such a simple scenario can at most provide the order of magnitude of the expected effect and cannot replace detailed calculations where e.g. the full geometry of the interaction process is taken into account.
The left part of Fig.~\ref{fig:skin} shows the predicted production probability $\sim$b$^{-1}dN_{\Py\PagL}/db$ for free $\PgL\PagL$ pairs (dashed histograms) and $\PgSm\PagL$ pairs (solid histograms) as a function of the impact parameter b for 1\GeV\ $\Pap$-$^{20}$Ne (red histograms)  and $\Pap$-$^{22}$Ne interactions (blue histograms).
The relative enhancement of $\PgSm\PagL$ and the suppression of $\PgL\PagL$ pairs in $^{22}$Ne with respect to $^{20}$Ne interactions is shown by the blue and red histogram in right part of Fig.~\ref{fig:skin}.
The $\PgL\PagL$ pair production is smaller by about 20\% in $^{22}$Ne compared to $^{20}$Ne for the whole impact parameter range. Such a value is in line with the qualitative considerations presented before.
The production of $\PgSm\PagL$ pairs shows a more complicated impact parameter dependence. While a suppression is observed at central collisions, the region of intermediate impact parameters shows similar yields for $^{20}$Ne and $^{22}$Ne. Going further into the periphery of the neon nuclei beyond b $>$3\,fm, the additional neutron skin of $^{22}$Ne enhances the $\PgSm\PagL$ production with respect to $^{20}$Ne considerably.
\begin{table}
 \centering
 \caption{Production yield of $\PgL\PagL$ and $\PgSm\PagL$ pairs in $\Pap$-Ne interactions. The number of produced pairs corresponds to 270 million inclusive interactions for each neon isotope. The last line gives the double-ratio for $\PagL\PgSm$ and $\PagL\PgL$ production.}
 \begin{tabular}{c r r}
\hline
Target            & \PgSm\PagL & \PgL\PagL\\ \hline\hline
$^{20}$Ne         & 3667     & 18808\\
$^{22}$Ne         & 4516     & 15733\\
ratio $^{22}$Ne/$^{20}$Ne& 1.23      & 0.84\\ \hline
ratio($\PgSm\PagL$)/ratio($\PgL\PagL$) & \multicolumn{2}{c}{1.46}\\ \hline
\end{tabular}
\label{tab:skin}
\end{table}
Of course, in $\Pap$-A collisions one can not constrain the impact parameter. In future studies we will also consider the co-planarity of the produced hyperon-antihyperon pair to extract additional information. However, in the present work we restrict the discussion to the total production ratios. Table \ref{tab:skin} gives the individual yields and the yield ratios for $^{22}$Ne and $^{20}$Ne targets.
Already the individual yields show a remarkable sensitivity to the additional neutron layer in $^{22}$Ne.
Since the yield ratios R for $\PgL\PagL$ and $\PgSm\PagL$ pairs are affected in opposite directions, the sensitivity is even more enhanced by forming the double ratio R($\PgL\PagL$)/R($\PgSm\PagL$). While in such double ratios systematic effects might partially cancel, further simulations are clearly needed to explore the model dependence of these novel observable.

\section{Approaching double hypernuclear spectroscopy at FAIR}
\label{sec:llhyp}
The confirmation of the substantial charge symmetry breaking in the J=0 ground states of the A=4 mirror hypernuclei $^4_{\PgL}$H and $^4_{\PgL}$He by precision measurements at MAMI \cite{Ess15} and at J-PARC \cite{Yam15} making use of novel techniques
demonstrates impressively the necessity to combine complementary methods in strangeness nuclear physics \cite{Gaz15}.
The case of double hypernuclei is another example for the need for a cooperation. Complex hypernuclear systems incorporating two (and more) hyperons can be created at J-PARC using kaon beams, in antiproton-nucleus interactions in $\panda$ at FAIR, in massive nucleus-nucleus collisions \cite{Bot07,Bot13,Bot15} in the CBM and NUSTAR experiments at FAIR, STAR at RHIC \cite{Sta10} and ALICE at CERN \cite{Ali15}. Experiments at J-PARC using kaon beams and nuclear emulsions will provide information on the absolute ground state masses of double hypernuclei. Two-particle correlation studies between single hypernuclei and $\PgL$ hyperons – similar to conventional two particle correlation studies in heavy ion reactions (see e.g. \cite{Poc87}) – may explore particle-unstable resonances in
double hypernuclei. The spectrum of excited particle stable states will be explored at the $\panda$ experiment by performing high resolution $\gamma$-spectroscopy. Complemented by hyperon-hyperon correlation studies in heavy ion collisions, these measurements will provide comprehensive information on the hyperon-hyperon interaction and the role of $\PgL\PgL$ - $\PgS\PgS$ - $\PgX N$ mixing in nuclei.
\begin{figure}[tb]
\includegraphics[width=0.55\textwidth]{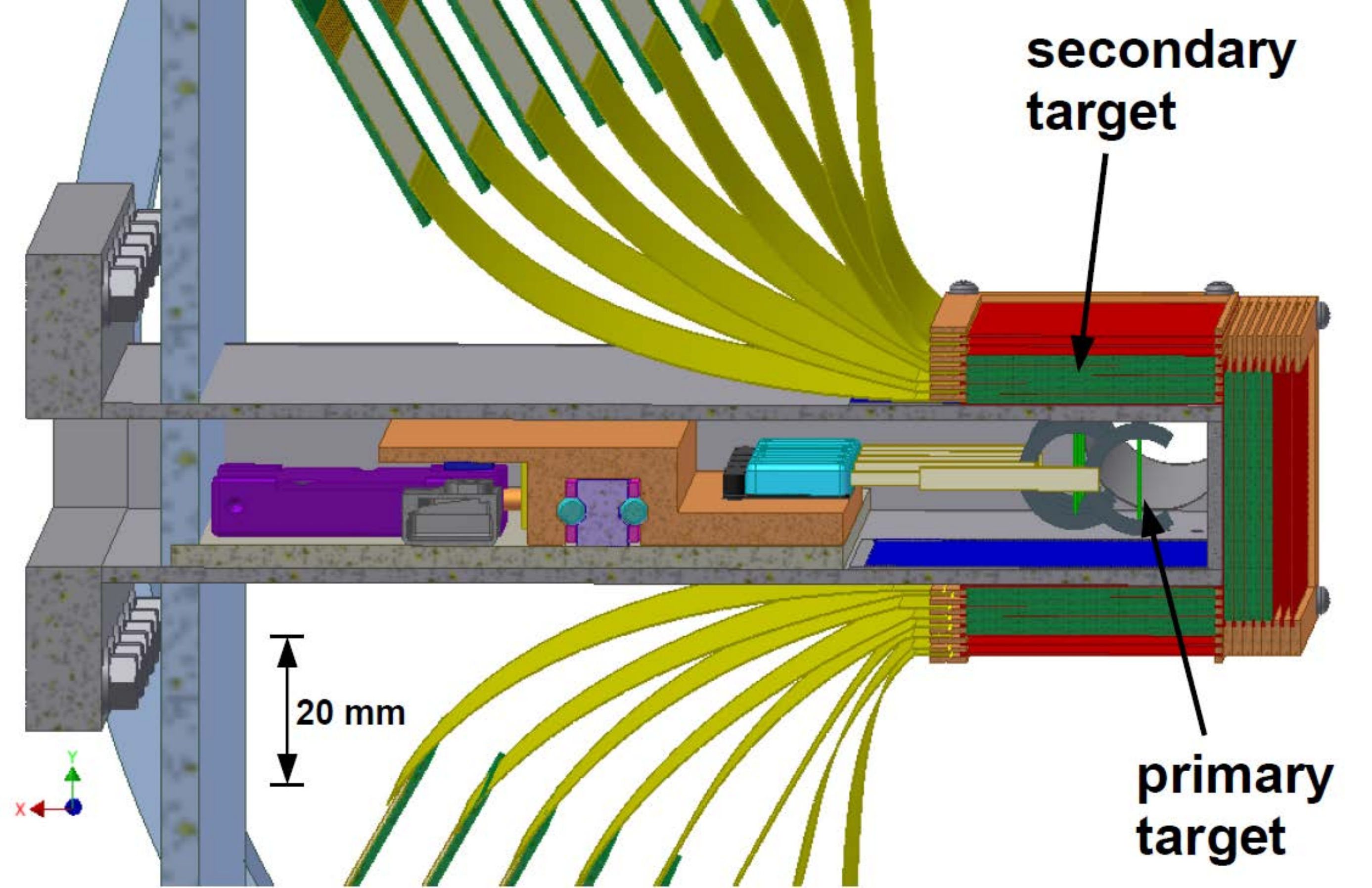}
\includegraphics[width=0.44\textwidth ]{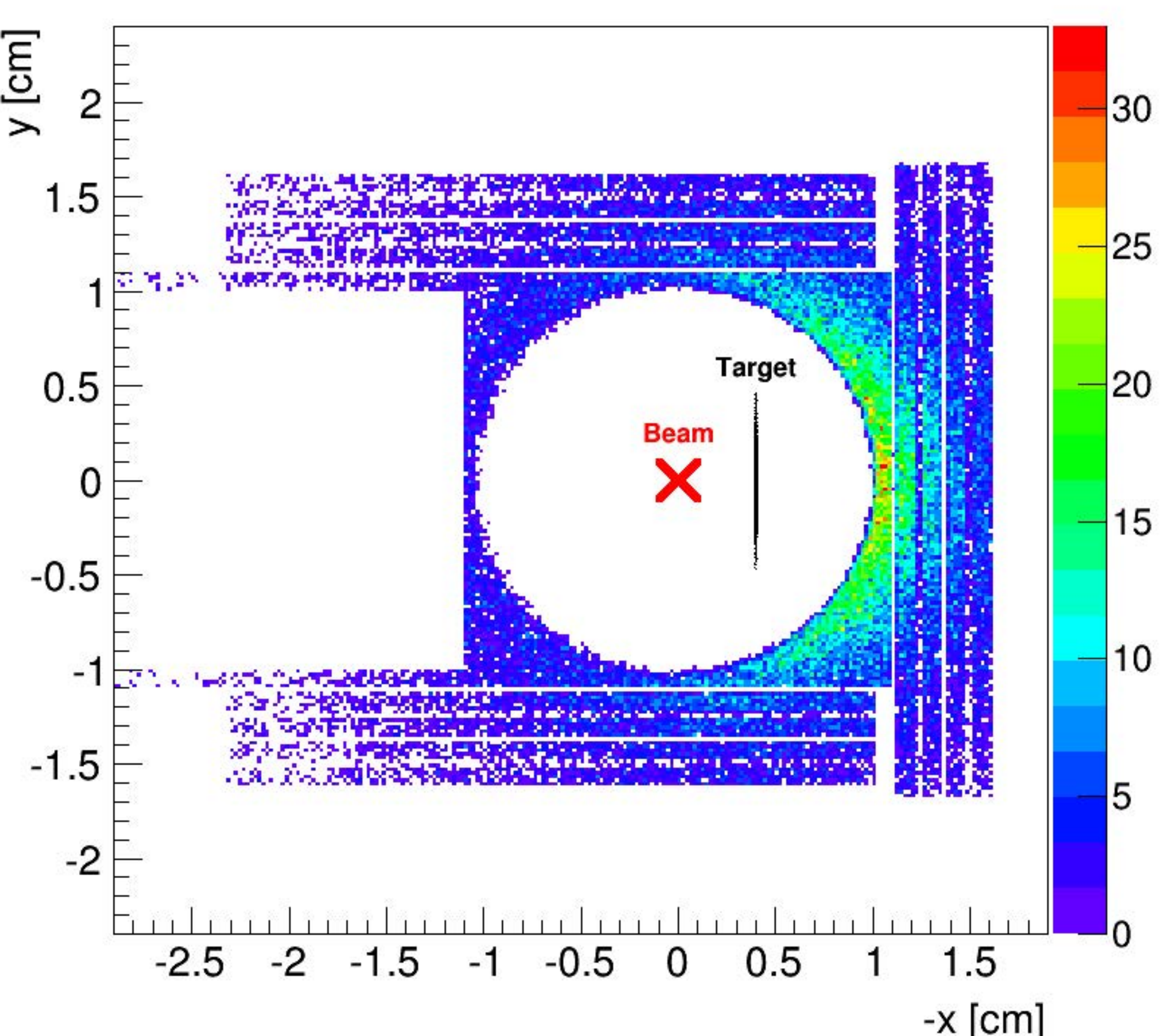}
\caption{Left: CAD drawing of the primary and secondary target of the setup. Right: Distribution of the $\PgXm$ stopping points in layers of the secondary target material in a plane transverse to the beam direction. Because of the short lifetime of $\PgXm$, a minimal distance between the primary target and the absorber material is essential to reach the optimal stopping probability.}
\label{fig:double1}
\end{figure}

\begin{figure}[b]
\includegraphics[width=0.60\textwidth]{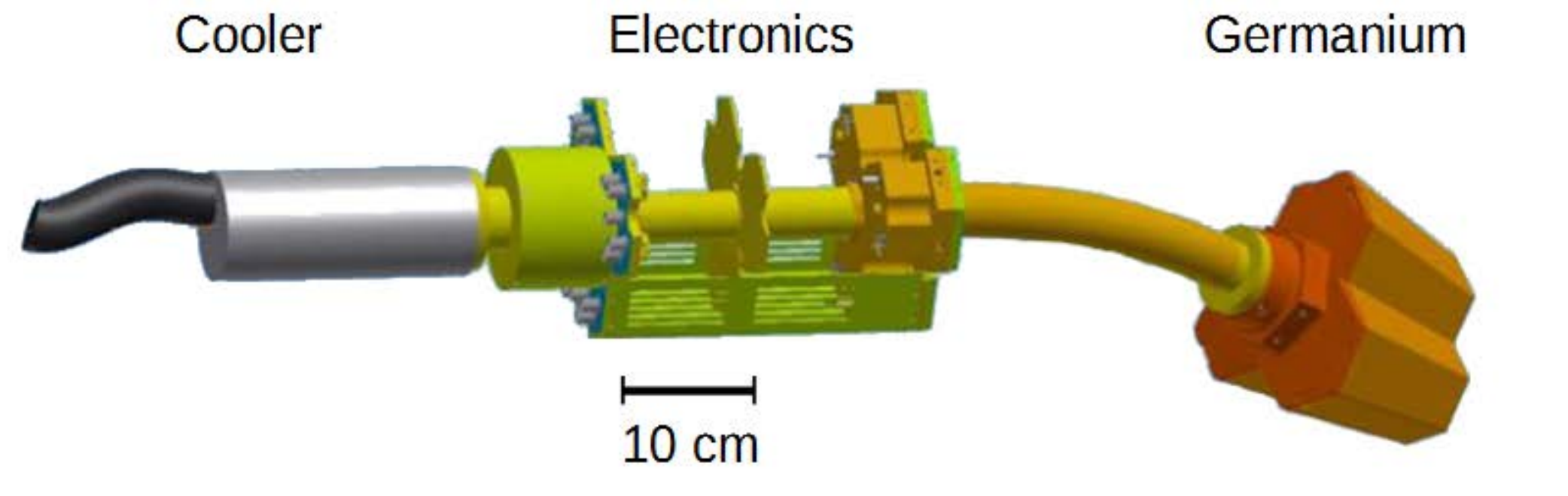}
\includegraphics[width=0.40\textwidth ]{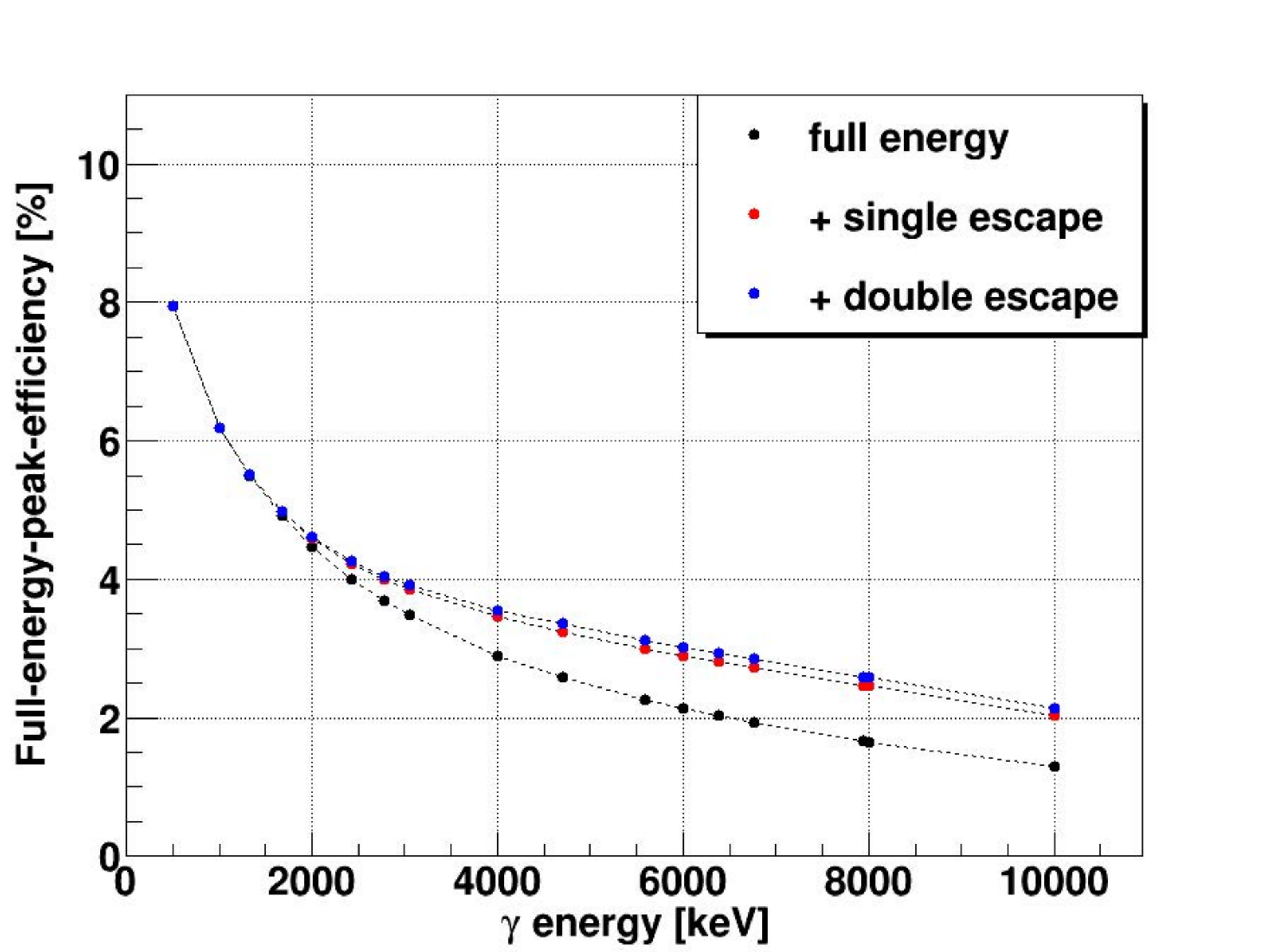}
\caption{Left: Final design of one triple of the $\panda$ Germanium Assembly PANGEA. Right: Expected full
-energy-peak efficiency of the PANGEA setup in $\panda$.}
\label{fig:double2}
\end{figure}

Since the first ideas of an antiproton storage ring at FAIR, the high resolution $\gamma$-spectroscopy of double hypernculei
is part of the core programme of the $\panda$ experiment \cite{Poc04,Pan09,San11}. To produce double hypernuclei in a more `controlled' way the conversion of a captured $\Xi^-$ and a proton into two $\Lambda$ particles can be used.
Fortunately, relatively low momentum $\Xi^-$ can also be produced using antiproton beams in $\Pap\Pp \rightarrow \PgXm \PagXp$ or $\Pap\Pn \rightarrow \PgXm \PagXn$ reactions happen in a complex nucleus where the produced $\Xi^-$ can re-scatter \cite{Poc04}. The advantage as compared to the kaon induced $\Xi$ production is that antiprotons are stable and can be retained in a storage ring thus allowing rather high luminosities.
Because of the two-step production mechanism, spectroscopic studies based on two-body kinematics cannot be performed for $\Lambda\Lambda$-hypernuclei and spectroscopic information can only be obtained via their decay products. The kinetic energies of weak decay products are sensitive to the binding energies of the two $\Lambda$ hyperons. While the double pionic decay of light double hypernuclei can be used as an effective filter to reduce the background, the unique identification of hypernuclei ground states only via their pionic decay is usually hampered by the limited resolution.
In addition to the general purpose $\panda$ setup, the hypernuclear experiment requires a dedicated primary target to produce low momentum {\PgXm}, an active secondary target of silicon layers and absorber material to stop the $\PgXm$ hyperons and to detect pions from the weak decay of hypernuclei, and a high purity germanium (HPGe) array as $\gamma$-detectors. The design of the setup and the development of these detectors is progressing (Figs.~\ref{fig:double1} and \ref{fig:double2}).

The primary target will consist of a diamond filament which will be moved in the halo of the antiproton beam to reach a constant luminosity during the measuring periods. Because of the short lifetime of the $\PgXm$ hyperons and their finite stopping time in the secondary target, it is essential to place the secondary absorbers as close as possible to the primary target to reach a maximum stopping probability. As a consequence the walls of the vacuum chamber close to the secondary target will be made of the absorber material. The right part in Fig.~\ref{fig:double2} shows the distribution of stopped {\PgXm} in case of boron as absorber material.

The PANGEA hodoscope made of 16 triple detectors (left part in Fig.~\ref{fig:double2}) will be mounted at backward angles. The right part of Fig.~\ref{fig:double2} shows the expected efficiency of this setup in $\panda$. At an average antiproton interaction rate of 5$\cdot$10$^6$\,s$^{-1}$ and with the present design, $\panda$ will produce approximately 33000 $\PgXm$'s per day stopped in boron absorbers of the secondary. Gating on the detection of two successive weak pionic decays this will provide approximately 10 detected $\gamma$-transitions per month for a specific $\PgL\PgL$-nucleus.

\section{Hyperatoms - doorway to s=-3 nuclear physics}
\label{sec:atoms}
\begin{figure}[b]
\includegraphics[width=0.37\textwidth]{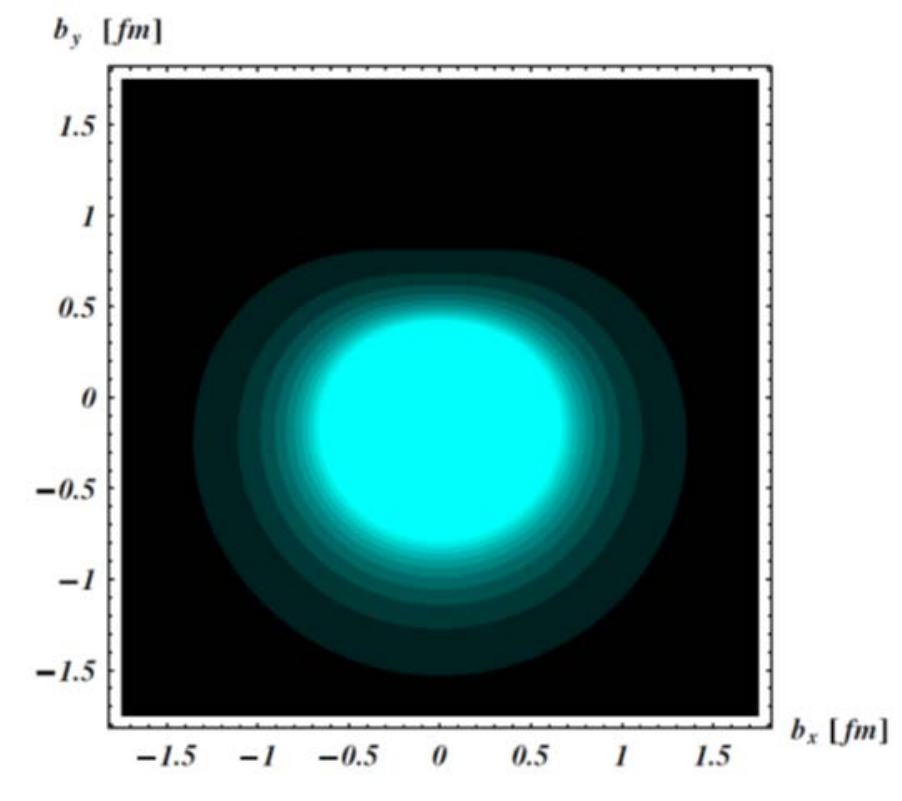}
\includegraphics[width=0.63\textwidth]{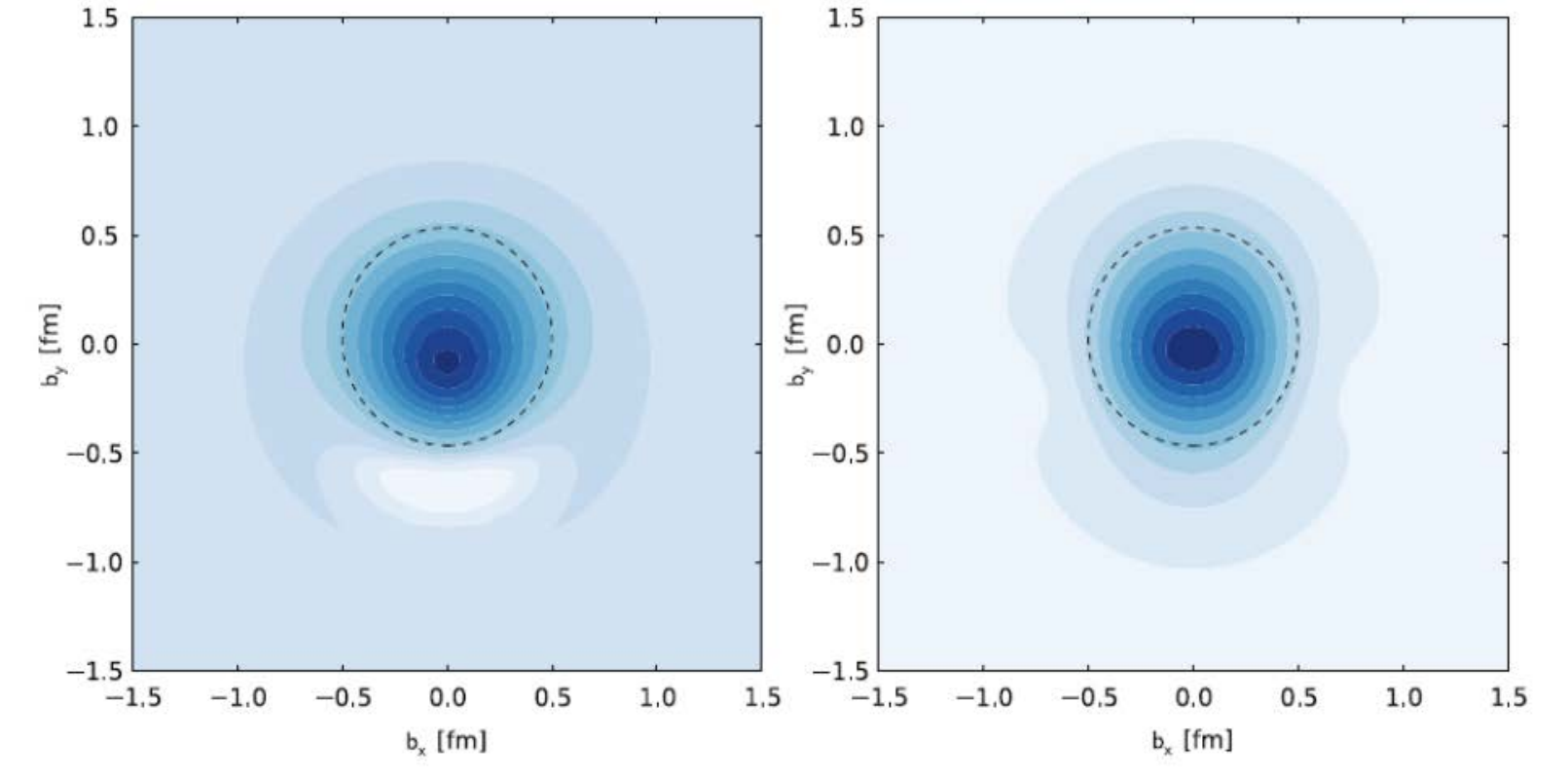}
\caption{Transverse charge densities of the proton (left, from Ref. \cite{Car08}) and the $\PgOm$  (from Ref. \cite{Ale10}) illustrating the intrinsic deformation of the baryons. In both cases the baryons are polarized along x axis.}
\label{fig:omega1}
\end{figure}

Like all composite particle, baryons are expected to be deformed objects (see Fig. \ref{fig:omega1}). However, for spin J=0 and 1/2 hadrons, the spectroscopic quadrupole $Q$ moment vanishes even though the intrinsic quadrupole moment $Q_0$ may be finite. On the other hand, for spin-3/2 particles the intrinsic quadrupole moment can be deduced from the spectroscopic moment according to (see e.g. \cite{Mes65})
\begin{equation}
Q=\frac{J(2J-1)}{(J+1)(2J+3)}Q_0.
\end{equation}
The long lifetime and its spin 3/2 makes the $\PgOm$ the only candidate to
obtain direct experimental information on the shape of an individual baryon.
A finite electric quadrupole form factors of spin-3/2 baryons would imply a deviation of the
baryon shape from spherical symmetry.
Also from the theoretical side the $\PgOm$ is particularly interesting because the valence quark content of $\PgOm$ is restricted to relative heavy strange quarks.
Unlike in the case of the nucleon, pion exchange is not relevant and the role of heavier mesons is strongly suppressed.  Therefore, meson cloud corrections to the valence quark
core are expected to be small \cite{Ram11}.
For negatively charged baryons, a positive (negative) quadrupole form factor
would signal an oblate (prolate) distribution of the three s-quarks.
In this sense, this measurement would be an important complement to the world wide activities trying to nail down the shape of the proton (see Fig.~\ref{fig:omega1}).

Measuring the quadrupole moment of the $\PgOm$, or setting a limit to its value, would provide very useful constraints on the composite models of baryons (see Tab. \ref{tab:qom}). Because contributions from light quarks are small, the
quadrupole moment of the $\PgOm$ will be a sensitive benchmark test for lattice QCD simulations.
Indeed, already more than 40 years ago Sternheimer and Goldhaber suggested to measure the electric
quadrupole Q$_\PgO$ of the $\PgOm$-baryon by the hyperfine splitting of levels in $\PgOm$-atoms \cite{Ste73}. However, the assumed large value for Q$_\PgO \approx$2\,$e\cdot$fm$2$ could not be supported by first theoretical estimates by Gershten and Zimov'ev within a nonrelativistic quark model \cite{Ger81}. Also all later calculations predicted an intrinsic quadrupole moment Q$_\PgO$ of the order of 0.01\,$e\cdot$fm$2$ (see Tab. \ref{tab:qom}).

\begin{table}
\caption{Predictions for the quadrupole moment of the $\PgOm$ baryon.}
 \centering
 \begin{tabular}{l r c}
  \hline
  Model             & Q$_\PgO$ [$e\cdot$fm$^2$] & Ref.\\ \hline\hline
  NRQM              & 0.02      & \cite{Ger81}\\
  NRQM              & 0.004     & \cite{Ric82}\\
  NRQM              & 0.031     & \cite{Isg82}\\
  SU(3) Bag model   & 0.052     & \cite{Kri87}\\
  NRQM with mesons  & 0.0057    & \cite{Leo90}\\
  NQM               & 0.028     & \cite{Kri91}\\
  Lattice QCD       & 0.0.005   & \cite{Lei92}\\
  HB$\chi$PT        & 0.009     & \cite{But94}\\
  Skyrme            & 0.024     & \cite{Kro94}\\
  Skyrme            & 0.0       & \cite{Yoo95}\\
  QM                & 0.022     & \cite{Buc97}\\
  $\chi$QM          & 0.026     & \cite{Wag00}\\
  GP QCD            & 0.024     & \cite{Buc02}\\
  QCD-SR            & 0.1       & \cite{Azi09}\\
  $\chi$PT+qlQCD    & 0.0086    & \cite{Gen09}\\
  Lattice QCD       & 0.0096$\pm$0.0002 & \cite{Ram11}\\
 \hline \hline
 \end{tabular}
 \label{tab:qom}
\end{table}

\begin{figure}[b]
\includegraphics[width=0.49\textwidth]{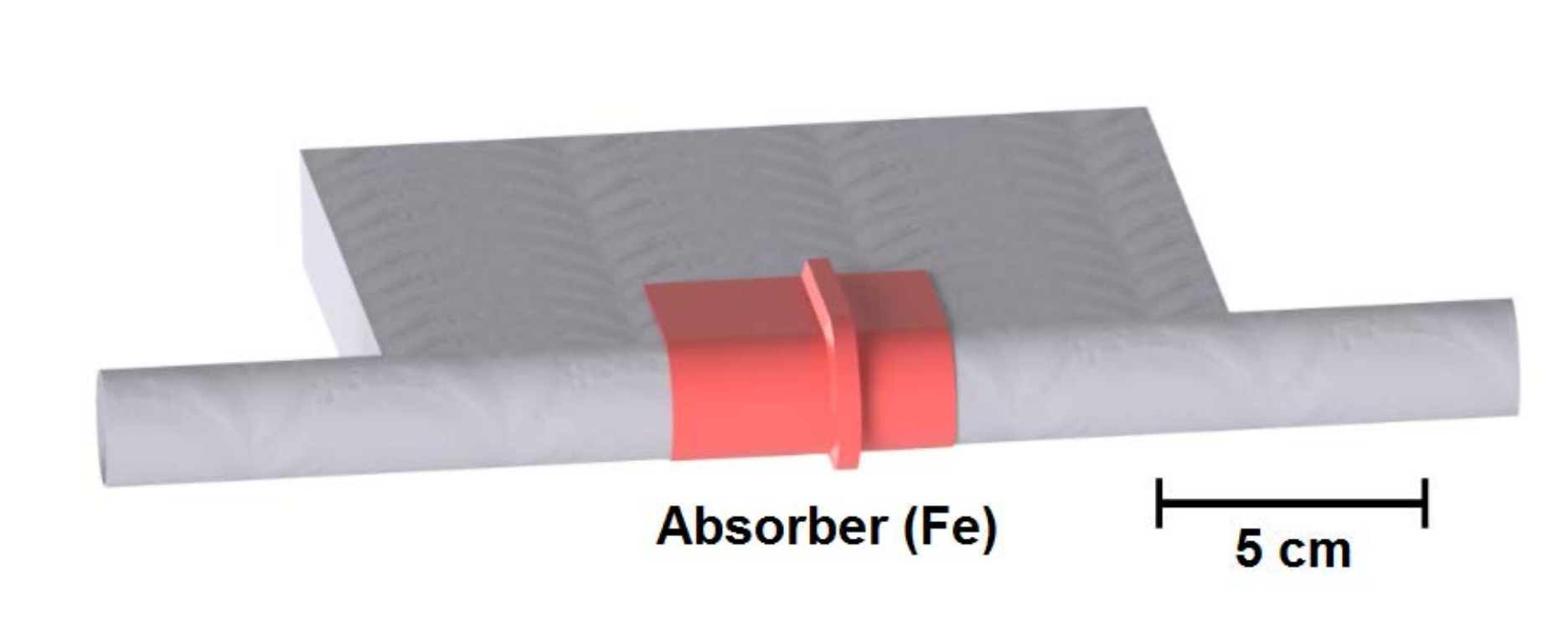}
\includegraphics[width=0.49\textwidth]{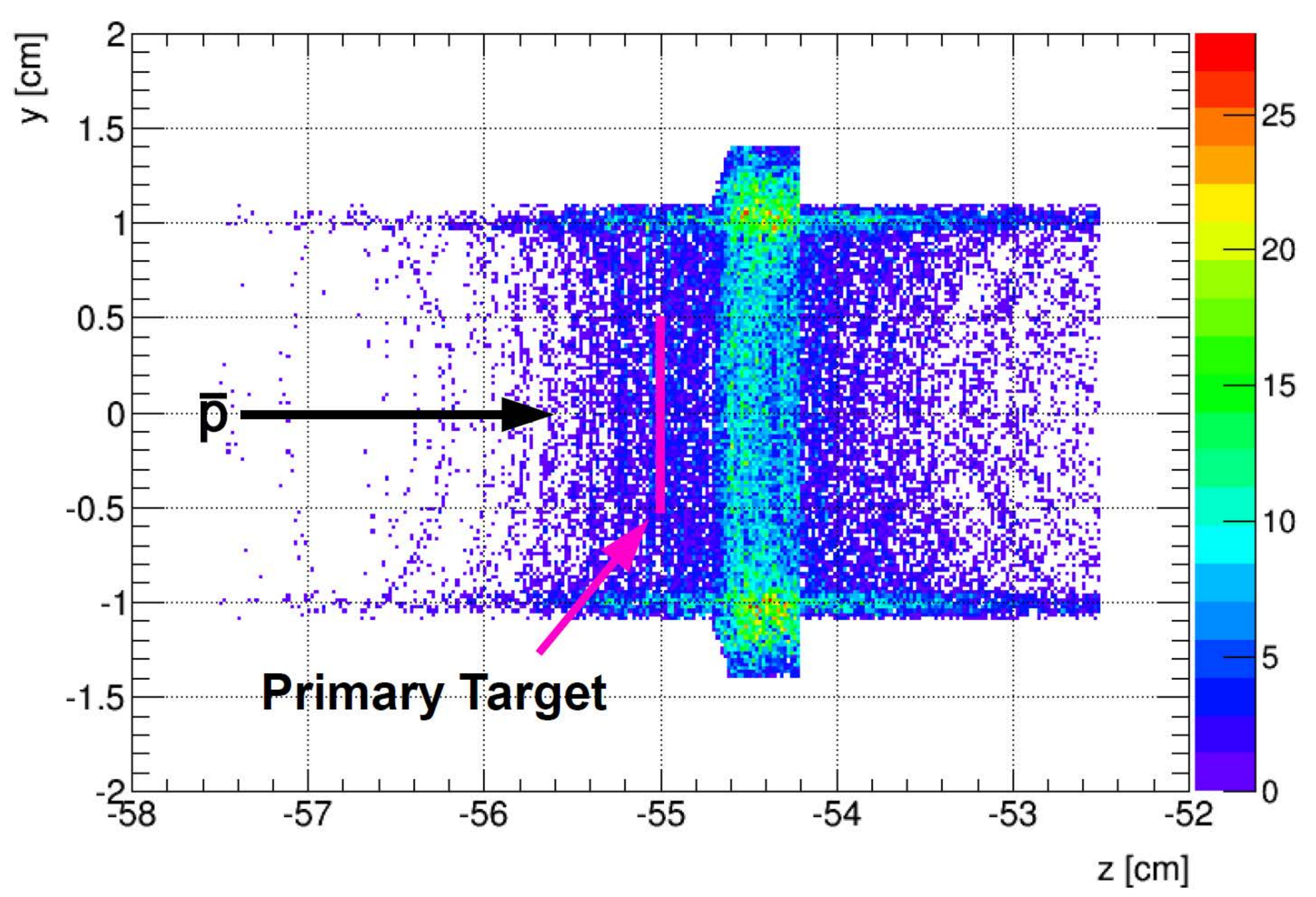}
\caption{Left: Schematic drawing of the secondary target for the hyperatom study at $\panda$.
Right: Stopping points predicted by full GEANT simulations which are based on GiBUU events. The shape of the rim is optimized for
maximum $\PgXm$ stopping and minimal losses of {$\gamma$}'s emitted from the hyperatoms.}
\label{fig:omega2}
\end{figure}
It is important to note that the deformation of the $\PgOm$ baryon is only one aspect of hyperatoms at $\panda$. The shift and broadening of transitions between orbits close to the nucleus provide a complementary tool for studying strong interactions and nuclear medium effects \cite{Bat99,Fri07}. Thus, $\PgOm$-hyperatoms represent a unique chance to explore the interaction of $|s|$=3 baryons in a nuclear system.

The difficulties in producing $\PgOm$-atoms and the high precision required for the $\gamma$-detection lead
to the sceptical conclusion made by
Batty 20 years ago \cite{Bat95}: \begin{quote}{\em ''The precision measurements of X-rays from {\PgOm}-Pb atoms will certainly require a future generation of accelerators and probably also physicists}.''\end{quote}
As shown by Alvarez \cite{Alv73}, three emulsion events observed in 1954 \cite{Eis54,Fry55} can be interpreted as $\PgOm$ decays (10 years prior to its discovery at Brookhaven\cite{Bar64}). Out of these 3 events, two can be attributed to the decay of atomically bound $\PgOm$.
This observation suggests that the formation of $\PgOm$-atoms is possible and may not be unusual once a $\PgOm$ hyperon has been slowed down.
Unfortunately, not even the elementary production cross section for ${\PgOm}{\PagO}$$^+$ pairs in antiproton-proton collisions is experimentally known (cf. Fig.~\ref{fig:xsec}) and even predictions are scarce \cite{Kai94} and may have large uncertainties. Therefore, quantitative predictions for the yield of atomic transitions in $\PgOm$-atoms are not possible at the moment. Assuming again an average antiproton interaction rate of 5$\cdot$10$^6$\,s$^{-1}$, a cross section for ${\PgOm}{\PagOp}$ pairs which is a factor of 100 lower compared to ${\PgXm}{\PagXp}$ \cite{Kai94} and a stopping probability for ${\PgOm}$ which is also a factor 100 lower than for ${\PgXm}$, $\panda$ might produce of the order of 5 $\PgOm$-atoms per day.

These considerations indicates that the study of $\PgOm$-atoms will not be a day-1 experiment. A well understood detection system and high luminosities will be mandatory for this measurement. As an intermediate step we therefore plan to study $\PgXm$-atoms at $\panda$. Even at a limited luminosity during the start-up phase, the large yield of stopped $\PgXm$ hyperons will produce a rate which is comparable to the rates expected at J-PARC. At the same time such a measurement will allow to develop and test the hypernuclear setup of $\panda$ under real running conditions.

Like in the case of the double hypernucleus study, a close proximity between the primary  target and secondary absorber is mandatory. In this case the absorbers can be heavy elements like e.g. Fe or Ta. As before, the vacuum chamber can be formed from this absorber material, thus optimizing the hyperon stopping probability.
At the same time the geometry of the secondary absorber should minimize the absorption of the atomic X-rays.
A first preliminary design of the secondary absorber is shown in Fig.~\ref{fig:omega2}. The shape of the rim is optimized for
maximum $\PgXm$ stopping at minimal losses of $\gamma$'s emitted from the hyperatoms.
The distribution of the $\PgXm$ stopping points are shown in the right part of Fig.~\ref{fig:omega2}. It is clear, that the final design of the secondary absorber
should be finalized by experimental data on the angular and momentum distributions of $\PgXm$ as well as $\PgOm$ hyperons.

To summarize, stored antiprotons beams in the GeV range represent a unparalleled factory for hyperon-antihyperon pairs. Their outstanding large production probability in antiproton collisions will open the floodgates for a series of new studies of strange hadronic systems with unprecedented precision.
Several of these unique experiments are possible at reduced luminosities in the commissioning phase of $\panda$, like the study of antihyperons in nuclear systems and the exclusive production of $\PgL\PagL$ and $\PgSm\PagL$ pairs in antiproton-nucleus interactions probing the neutron and proton distributions. Also, the high resolution spectroscopy of multistrange $\Xi$-atoms will already be feasible during the initial phase. The high resolution spectroscopy of $\PgL\PgL$-hypernuclei will require an interaction rate in the region of 5$\cdot$10$^6$. The spectroscopy of $\PgOm$-atoms will be challenging, but seems possible. Looking even further into the future, an $HESR-\overline{HESR}$ antiproton-proton collider could revolutionize the field of strangeness nuclear physics by producing momentum and polarization tagged hyperon beams with extremely low momenta.

This work is partially contained in the PhD theses of Sebastian Bleser and Marcell Steinen.








\end{document}